\documentclass[12pt]{article}
\usepackage{times}
\usepackage{geometry}
\usepackage{floatflt}
\usepackage[utf8]{inputenc}
\usepackage{enumitem,amssymb}
\usepackage{ragged2e}
\newlist{thematic}{itemize}{8}
\setlist[thematic]{label=$\square$}
\usepackage{pifont}
\newcommand{\cmark}{\ding{51}}%
\newcommand{\done}{\rlap{$\square$}{\raisebox{2pt}{\large\hspace{1pt}\cmark}}%
\hspace{-2.5pt}}

\usepackage{jheppub}   
\usepackage[dvipsnames,svgnames,x11names]{xcolor}
\usepackage{sectsty}
\usepackage[outercaption]{sidecap}
\sidecaptionvpos{figure}{c}

\newcommand\chandra{{\sl Chandra}}

\newcommand\xrism{{\sl XRISM}}
\newcommand\athena{{\sl Athena}}
\newcommand\lynx{{\sl Lynx}}

\newcommand{\bigquestion}[1]{{\bf  \color{purple} #1}}

\newcommand{\arcmin}{\hbox{$^\prime$}}               
\newcommand{\arcsec}{\hbox{$^{\prime\prime}$}}       

\definecolor{DarkGreen}{rgb}{0.0, 0.3, 0.0}
\definecolor{purple}{rgb}{0.5, 0.0, 0.5}
\definecolor{red}{rgb}{1, 0.0, 0.0}
\definecolor{green}{rgb}{0, 1.0, 0.0}

\newcommand{\Mvir}{\mathrel{M_{\rm vir}}}

\newcommand{\Msol}{\mathrel{M_\odot}}

\newcommand{\taue}{\mbox{$\tau_{\mbox{\tiny e}}$}}

\newcommand{\keV}{{{\rm keV}}}

\newcommand{\Te}{{T_{\rm e}}}
\newcommand{\kB}{\mbox{$k_{\mbox{\tiny B}}$}}

\newcommand{\vz}{\mbox{${v}_{\mbox{\bf \tiny z}}$}}

\newcommand{\Tcmb}{\mbox{$T_{\mbox{\tiny CMB}}$}}

\newcommand{\Moon}{$M_{Moon}$}                          

\newcommand{\apjl}{\rm ApJL}
\newcommand{\apjs}{\rm ApJS}

\newcommand{\aap}{\rm A$\&$A}

\def\ao{{\it Appl.~Opt.}}           

\def\3he{$^3{\rm He}$}
\def\arcdeg{\hbox{$^\circ$}}

\def\lsim{\mathrel{\lower2.5pt\vbox{\lineskip=0pt\baselineskip=0pt
           \hbox{$<$}\hbox{$\sim$}}}}

\def\gsim{\mathrel{\lower2.5pt\vbox{\lineskip=0pt\baselineskip=0pt
           \hbox{$>$}\hbox{$\sim$}}}}

\begin{document}
\raggedright
\huge
Astro2020 Science White Paper \linebreak

A High-resolution SZ View of the Warm-Hot Universe
\linebreak
\normalsize

\noindent \textbf{Thematic Areas:} \hspace*{60pt} $\square$ Planetary Systems \hspace*{10pt} $\square$ Star and Planet Formation \hspace*{20pt}\linebreak
$\square$ Formation and Evolution of Compact Objects \hspace*{31pt} $\done$ Cosmology and Fundamental Physics \linebreak
  $\square$  Stars and Stellar Evolution \hspace*{1pt} $\square$ Resolved Stellar Populations and their Environments \hspace*{40pt} \linebreak
  $\done$    Galaxy Evolution   \hspace*{45pt} $\square$             Multi-Messenger Astronomy and Astrophysics \hspace*{65pt} \linebreak
  
\textbf{Principal Authors:}

Names:	Tony Mroczkowski$^1$, Daisuke Nagai$^2$
 \linebreak						
Institutions:  1) European Southern Observatory, 2) Yale University
 \linebreak
Email:  tony.mroczkowski@eso.org, daisuke.nagai@yale.edu
 \linebreak
 Phone:  +49-89-32006174, +1-203-432-5370
 \linebreak
 
\textbf{Co-authors:} 
Paola Andreani (ESO),
Monique Arnaud (CEA),
James Bartlett (APC/U.Paris Diderot),
Nicholas Battaglia (Cornell), 
Kaustuv Basu (Bonn),
Esra Bulbul (CfA),
Jens Chluba (Manchester),
Eugene Churazov (MPA/IKI),
Claudia Cicone (INAF/Oslo),
Abigail Crites (Caltech),
Nat DeNigris (U.Mass),
Mark Devlin (U.Penn),
Luca Di Mascolo (MPA),
Simon Dicker (U.Penn),
Massimo Gaspari (Princeton), 
Sunil Golwala (Caltech),
Fabrizia Guglielmetti (ESO),
J.\ Colin Hill (IAS/CCA),
Pamela Klaassen (ROE),
Tetsu Kitayama (Toho),
R\"udiger Kneissl (ESO/ALMA),
Kotaro Kohno (U.Tokyo),
Eiichiro Komatsu (MPA),
Mark Lacy (NRAO),
Brian Mason (NRAO),
Kristina Nyland (NRC, resident at NRL),
Charles Romero (U.Penn),
Jack Sayers (Caltech),
Neelima Sehgal (Stony Brook),
Sara Simon (U.Michgan),
Rashid Sunyaev (MPA/IKI/IAS),
Grant Wilson (U.Mass),
Michael Zemcov (RIT),
John ZuHone (CfA)
\linebreak

\justify
\textbf{Abstract:}
The Sunyaev-Zeldovich (SZ) effect was first predicted nearly five decades ago, but has only recently become a mature tool for performing high resolution studies of the warm and hot ionized gas in and between galaxies, groups, and clusters. Galaxy groups and clusters are powerful probes of cosmology, and they also serve as hosts for roughly half of the galaxies in the Universe.
In this white paper, we outline the advances in our understanding of thermodynamic and kinematic properties of the warm-hot universe 
that can come in the next decade through spatially and spectrally resolved measurements of the SZ effects.  Many of these advances will be enabled through new/upcoming millimeter/submillimeter (mm/submm) instrumentation on existing facilities, but truly transformative advances will require construction of new facilities with larger fields of view and broad spectral coverage of the mm/submm bands.

\pagebreak

\section{Introduction}
\label{sec:intro}
\vspace{-2mm}
Structures such as galaxies, groups, clusters, and the cosmic web filaments that connect them contain a warm or hot ($>10^5$~K), gravitationally-bound component that dominates their baryon count.  In such structures -- from individual galaxies to galaxy clusters -- this gas informs us about the total mass, composition, accretion history, and role of AGN feedback, while 
filaments comprising the cosmic web are expected to host the majority of the so-called `missing baryons' at redshifts $z\lesssim 3$ (i.e.\ when the Universe was $\sim$20\% its present age) \citep{Fukugita1998,Cen1999,Bregman2009,Shull2012}.
Galaxy groups and clusters are powerful probes of cosmology, and they also serve as hosts for roughly half of the galaxies in the Universe.
As discussed here, a suite of secondary anisotropies in the cosmic microwave background (CMB), due to Thomson scattering by free electrons and known as the Sunyaev-Zeldovich (SZ) \citep{Zeldovich1969,SZ1972,Sunyaev1980} effects, can provide useful, redshift-independent probes of the ionized gas that makes up the dominant and optically invisible baryonic component of the Universe \citep[][for a recent review]{Mroczkowski2019}. 

\begin{SCfigure}[1.5][h!]
  \hspace{-4mm}
   \includegraphics[width=70mm, clip=true, trim=0 0 0 0]{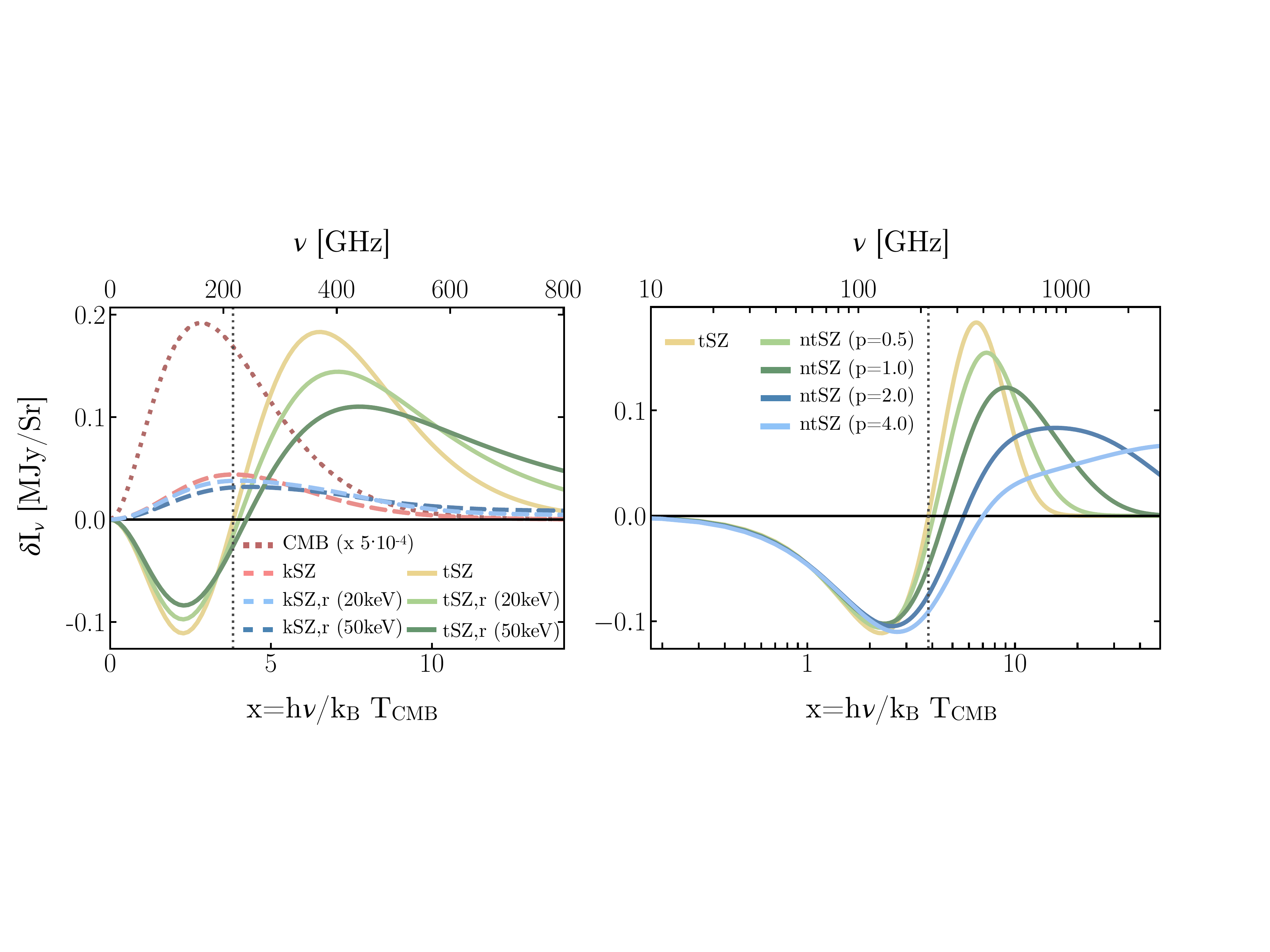}
   \caption{The tSZ (solid) and kSZ (dashed) effects, including relativistic corrections for various electron temperatures. 
The plots show the SZ effect distortions $\delta I_\nu$ for different scenarios and compositions as a function of (dimensionless) frequency $\nu$ ($x=h \nu / \kB \Tcmb$), assuming an optical depth $\taue=10^{-2}$, a Compton parameter $y=10^{-4}$, and a line-of-sight peculiar velocity
$\vz = +1000~\mathrm{km\,s^{-1}}$ (typical of a massive galaxy cluster). The dotted, dark red curve illustrates the Planckian shape of the primary CMB spectrum scaled by a factor of $5\times 10^{-4}$ for comparison purposes. 
	Figure from \cite{Mroczkowski2019}.
   }\label{fig:szspectrum}
\end{SCfigure}

In massive systems ($\Mvir \gtrsim 10^{13} \Msol$), the thermal SZ (tSZ) effect is typically strongest. 
The tSZ effect is proportional to the integrated thermal pressure of the gas along the line-of-sight, thus providing a direct calorimetric probe of the thermal energy content of the ionized gas in the intracluster, intragroup, circumgalactic, and warm-hot intergalactic media (ICM, IGrM, CGM \& WHIM).
Generally more challenging to observe is a Doppler shift in the primary CMB that measures the mass-weighted velocity of the gas along the line-of-sight known as the kinetic SZ (kSZ) effect. 
Further, the impact of relativistic corrections to the tSZ effect \citep[e.g.,][]{Itoh98, Chluba2012SZpack} 
can become significant for gas $\gtrsim 5~\keV$ \citep{Hurier2016rSZ, Erler2018}, providing an X-ray-independent probe of temperature in massive galaxy clusters out to the redshift of formation.
Fig.\ \ref{fig:szspectrum} shows the tSZ and kSZ effects and the impact of relativistic effects (rSZ) as a function of frequency $\nu$. 
In the coming decade, high-resolution, multi-frequency observations of the SZ effects will provide new insights into the thermodynamic and kinematic properties of the ICM, IGrM, CGM, and WHIM, especially in the low-density regions in the outskirts of dark matter halos and their connection to the cosmic web throughout the epochs of galaxy and cluster formation. 

\vspace{-3mm}
\vspace{-5mm}

\vspace{-6mm}
\section{Probing Thermal Structure and Evolution of the Universe}
\label{sec:agn}
\vspace{-3mm}
\noindent 
\bigquestion{What are the thermodynamic states of the ICM, IGrM, CGM, and WHIM? When do they first form, and how do they evolve? How do they impact star formation and galaxy evolution?} 

\smallskip
\noindent
{\bf  ICM Thermodynamics:} 
Measurements of the tSZ effect will provide important information about the thermodynamic structure of the ICM, including the impacts of feedback, bulk and turbulent motions, substructure, and cluster asphericity. Since the total thermal energy content is determined primarily by the gravitational potential of the cluster, the integrated tSZ effect signal, $Y_{\rm SZ}$, serves as a robust  proxy for total mass \citep[e.g.,][]{motl2005,Nagai2006,Battaglia2012,Kay2012,krause2012,Yu2015}. 
However, despite the robustness of $Y_{\rm SZ}$ as a mass proxy, there can be large deviations from self-similarity, particularly during ubiquitous cluster mergers - the most energetic events since the Big Bang \cite{Sarazin2002}. 
These cluster mergers can induce significant tSZ and kSZ effect substructure through compression \citep{poole2007,wik2008} and bulk and turbulent motion \citep{Ruan2013}. 

\begin{SCfigure}[1.4][h!]
\hspace{-3mm}
 \includegraphics[height=43mm]{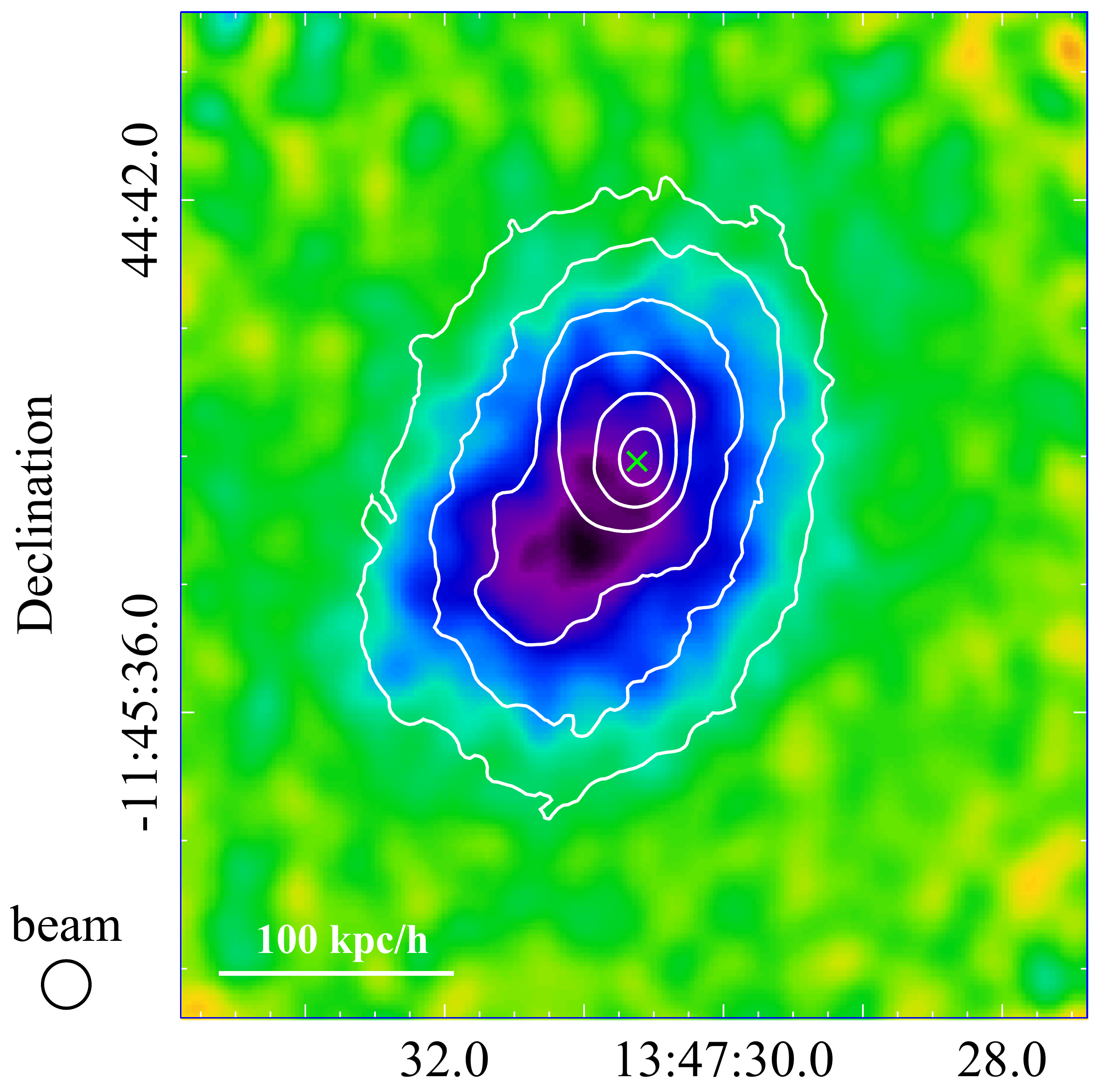}
 \includegraphics[height=43mm]{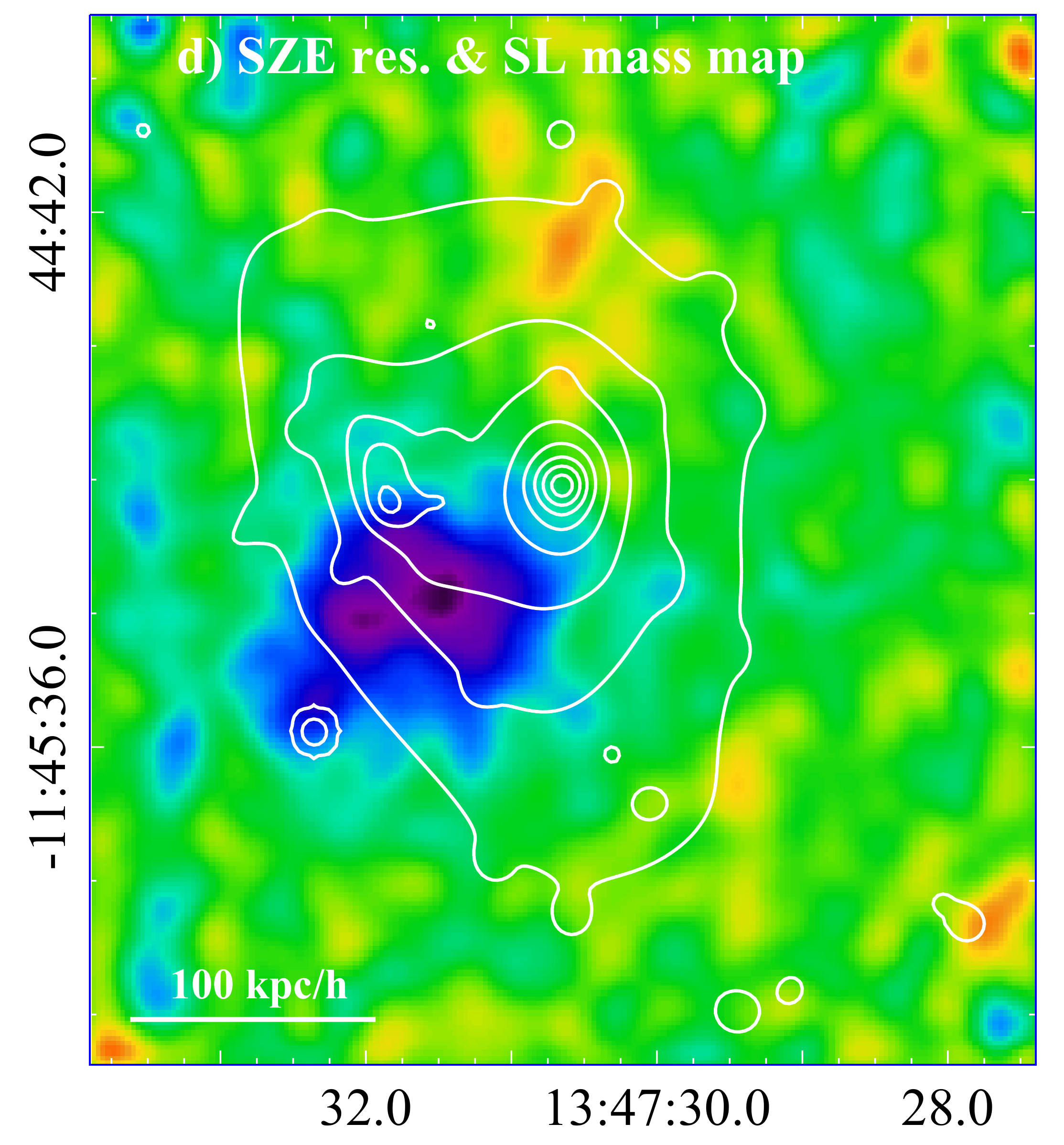}
\caption{\textit{Left:} ALMA+ACA image ($\sim$5\arcsec\ resolution) of RX~J1347.5-1145 with X-ray contours overlaid.
\textit{Right:} Substructure in the tSZ effect signal revealed after subtraction of a mean profile excluding the SE quadrant.  Lensing contours are overlaid, revealing the location of the dark matter component.
Figures from \cite{Ueda2018}.
}
\label{fig:rxj1347}
\end{SCfigure}

Thanks to the advances of high-angular-resolution imaging experiments, it is just now becoming possible to study the evolution of -- and characterize pressure substructures in -- the ICM through tSZ effect studies, assessing the impact on cluster cosmology. Substructures detected through the tSZ effect are related to gas compression driven by merger events associated with infalling substructure. Fig.~\ref{fig:rxj1347} shows one such example, revealing the offset between the peaks of the tSZ effect, X-ray, and strong lensing signals in the merging of a subcluster in RX~J1347.5-1145 \citep{Ueda2018}, each providing complementary views of the thermodynamic properties of ICM and the distribution and nature of dark matter (e.g.\ the self-interaction cross-section).

\vspace{-2mm}
\vspace{-4mm}

\noindent
{\bf Measuring Gas Temperature using relativistic corrections to the tSZ:}
In the near future, an exciting prospect for studying mergers and ICM substructure will also come from improved constraints on the rSZ. The rSZ can provide a redshift-independent direct measurement of the gas temperature, approximately weighted by electron pressure (n.b.\ this is independent of X-ray spectroscopy, which for comparison is weighted by the X-ray emission, which scales as density-squared). This topic is explored in more detail in the white paper by Basu \& Erler et al.

\begin{SCfigure}[1.4][h!]
 \hspace{-3mm}
 \includegraphics[height=48mm]{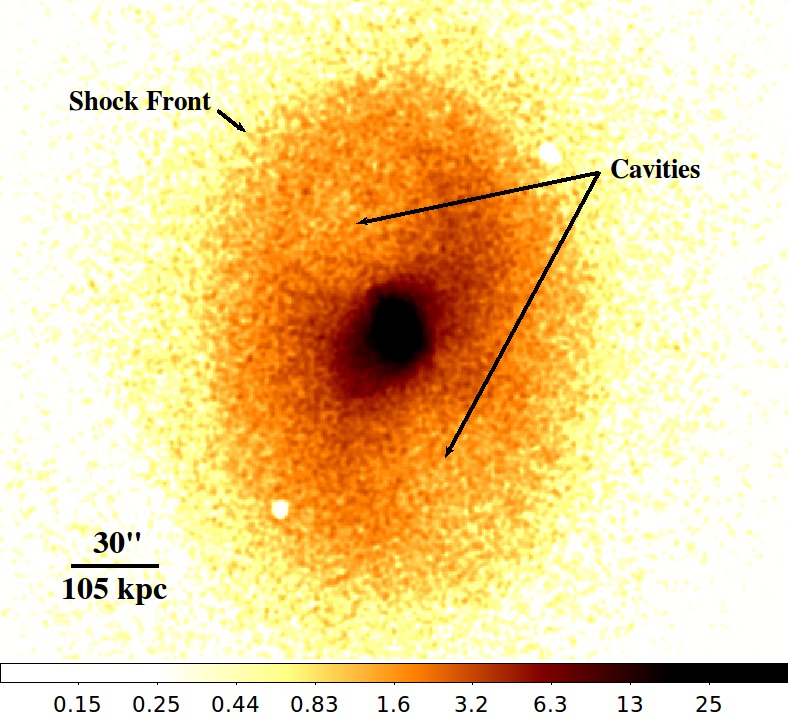}
 \hspace{-2mm}
 \includegraphics[height=48mm]{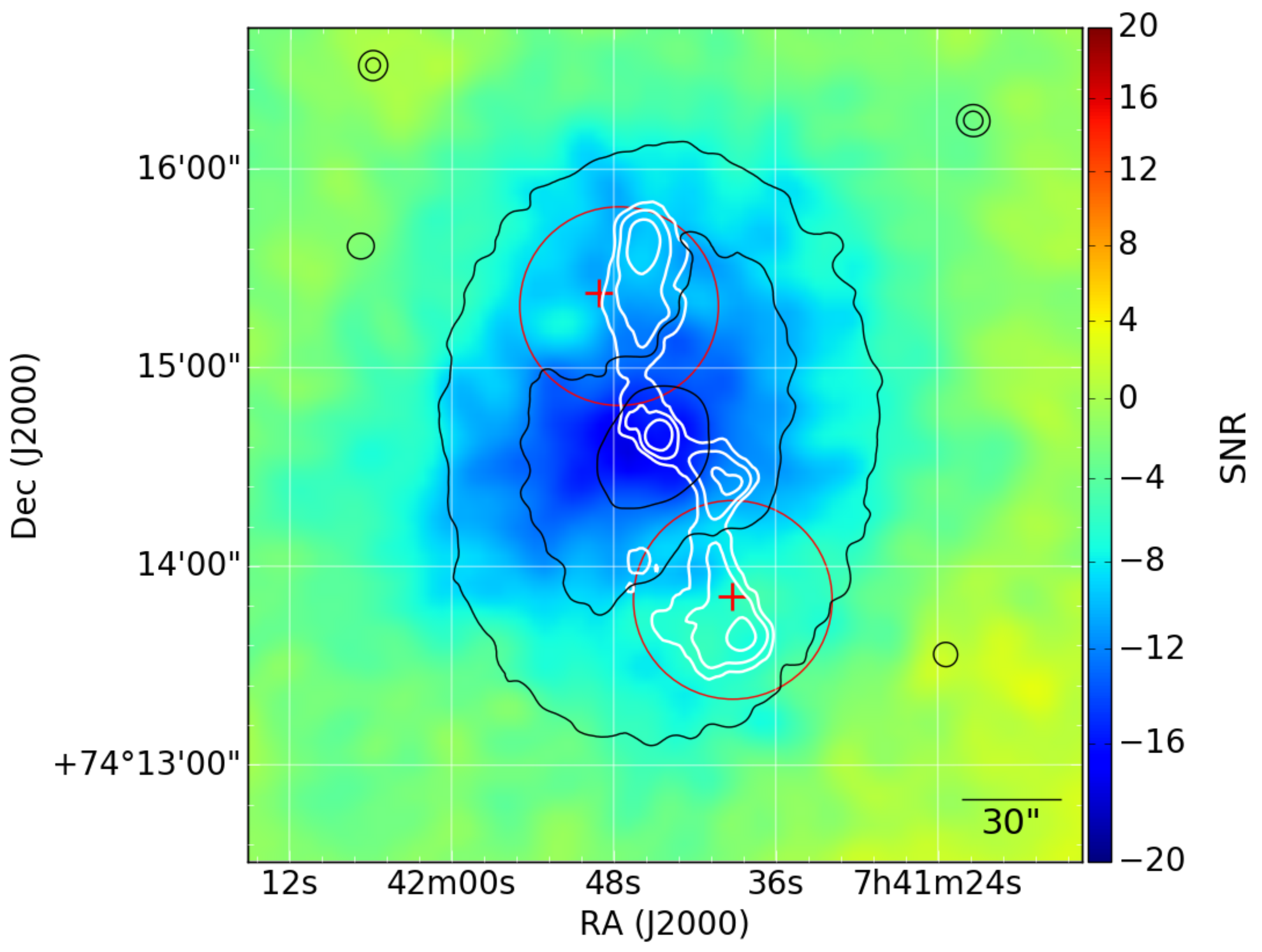}
	\caption{\chandra\ X-ray (left) and CARMA 30~GHz S/N (right) maps of MS0735.6.  White contours on right delineate the radio bubbles; black are from X-ray, showing the cavity positions. 
	Panels from \cite{Abdulla2018,Vantyghem2014}.}
\label{fig:ms0735}
\end{SCfigure}

\smallskip
\noindent
{\bf Constraining Models of AGN feedback:}
High-resolution, multi-frequency SZ effect observations also promise to provide new insights into the physics of feedback from supermassive black holes and its impact on the evolution of groups/clusters, and thus probe different mechanisms 
thought to suppress runaway cooling flows \citep[e.g.][]{Gaspari2011,Li2015,Yang2019}.
The right panel of Fig.~\ref{fig:ms0735} shows the recent CARMA 30 GHz measurements of the AGN-inflated radio bubbles (X-ray cavities, left) in MS0735.6+7421 ($z=0.21$), indicating the lack of a tSZ decrement from these features \citep{Abdulla2018}. 
Further, a recent ALMA study of the SZ effect from the hyperluminous quasar HE~0515-4414 constrain the ratio of the kinetic luminosity of the thermal wind (containing particles with energies $\sim$few keV) to the bolometric luminosity of the quasar to only $\sim 0.01$\% \citep{Lacy2019}, suggesting that thermal winds alone cannot be the dominant feedback mechanism in quasar host galaxies.

Although previous X-ray measurements indicate that cavities could in principle be supported by plasma that is too hot and diffuse to emit in the X-ray band \citep{mcnamara2007,Blanton2011,McNamara2012,Vantyghem2014}, the lack of a tSZ decrement from these cavities begins to constrain their composition and suggests that, if thermal, the gas should be at temperatures $>100$~keV.  Therefore, non-thermal cosmic-ray electrons and protons are likely the primary support mechanism \citep{Pfrommer2005,Pfrommer2007,prokhorov2012}. 
One exciting prospect is that by covering more of the SZ increment, one could directly probe non-thermal deviations from the SZ effect (see e.g.\ \cite{Pfrommer2005,Colafrancesco2006,Prokhorov2012rSZ,Mroczkowski2019}).
Unraveling the nature of AGN feedback thus drives a few key instrumentation requirements: 1) higher sensitivity, as most bubbles have enthalpies (and SZ signals) 2-4 orders of magnitude lower than those in MS0735.6, 2) broad spectral coverage of the mm/submm bands, and 3) sufficient spatial dynamic range to resolve more typical (few arcsec) bubbles while recovering the bulk SZ effect signal from the entire cluster (generally on scales of 10\arcmin--30\arcmin).

\vspace{-4mm}
\vspace{-4mm}

\smallskip
\noindent
{\bf Testing Galaxy Formation Models with CGM observations:}
From galaxies to galaxy clusters, the effects of feedback leave an imprint on their tSZ effect profiles, and should be measurable using large samples \citep[e.g.,][]{scannapieco2004,granato2004,croton2006,thacker2006,dimatteo2008}. Feedback effects are particularly strong in the inner regions of lower mass systems, and have been observed by stacking large samples of clusters, groups, and even galaxies \citep{Greco2015}. 
Such measurements, if extended to lower mass regimes, will strongly constrain the subgrid models used to describe galaxy formation and AGN feedback in cosmological simulations (see the white paper by Battaglia \& Hill et al.). \emph{Ultimately, the stacking measurements of today will become the radially-averaged and direct imaging measurements of tomorrow.}

\smallskip
\noindent
{\bf Probing the Missing Baryons and Low-Overdensity Environments:}
Since the tSZ effect depends linearly on density (c.f. X-ray surface brightness, which drops as density-squared), it has long been promoted as a tool for probing the low density gas in and accreting from cosmic web filaments.  
These regions are particularly important for revealing the environmental properties that drive the quenching of star formation in galaxies \cite{Zinger2018}.
Recent works have begun to fulfill this, putatively detecting the WHIM \citep{deGraaff2017,Kovacs2018,Tanimura2019} and virial shock in a cluster's outskirts \citep{Hurier2019}.  
While current subarcminute-resolution instruments are limited to 1\arcmin--6\arcmin\ scales, in order for this avenue of research to advance to direct imaging measurements, large fields of view probing out to and beyond the virial radius ($\sim$ few Mpc h$^{-1}$, typically 10\arcmin--30\arcmin\ at intermediate and high redshift) at high spatial dynamic range are required.  In such low temperature regions, the kSZ effect can be expected to play an increasingly important role (see the white paper by Battaglia \& Hill et al.\ \& \citep{Hill2016,Planck2016_XXXVII,Schaan2016,Lim2017}).

\vspace{-6mm}
\section{Shocks, Mergers \& Turbulence in the Growth of Structure}
\vspace{-3mm}
\noindent
\bigquestion{What are the kinematic properties of ICM, IGrM, and CGM plasma? What is the level of bulk and turbulent gas motions and how do they impact hydrostatic mass estimates?}

\begin{SCfigure}[1.6][h!]
 \hspace{-0mm}
    \includegraphics[trim=2.5cm 0cm 0cm 0cm, height=43mm]{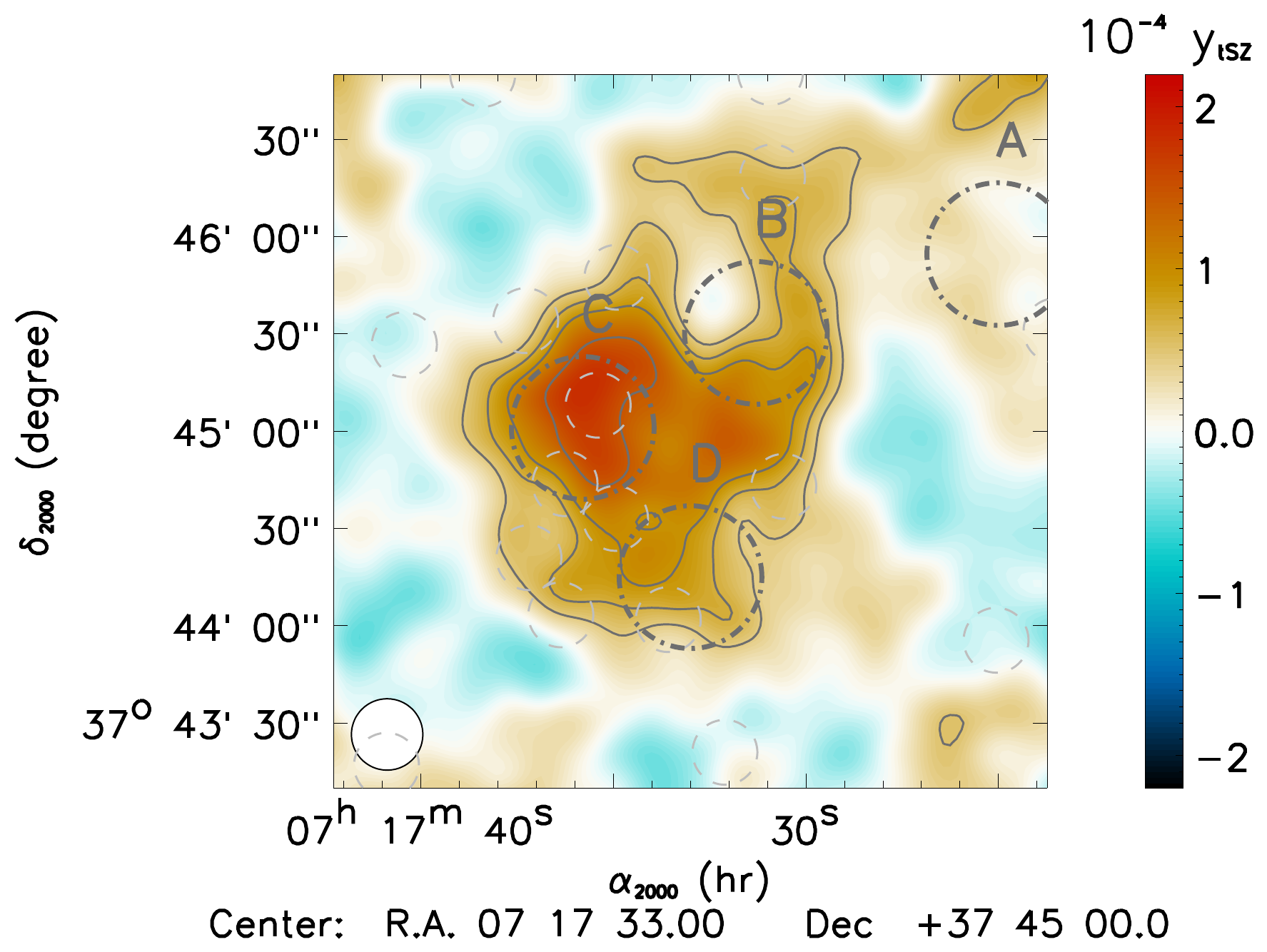}
     \hspace{-2mm}
    \includegraphics[trim=4.55cm 0cm 0cm 0cm, clip=true, height=43mm]{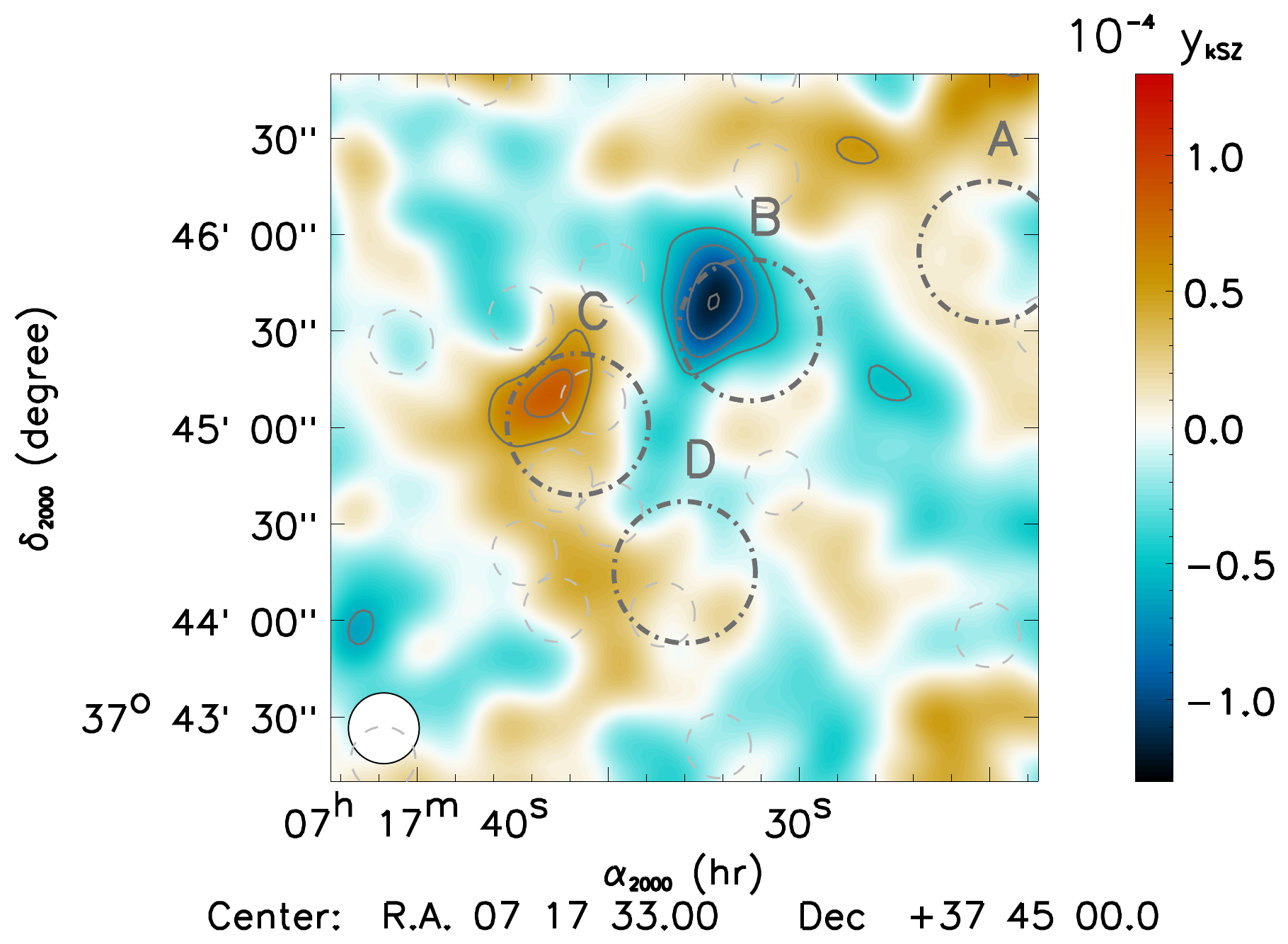}
   \caption{NIKA constraints on MACS~J0717.5, using 150 \& 260 GHz to reconstruct the tSZ (left) and kSZ (right) effect signals and to clean for contaminants.   The tSZ effect map reveals $\Te\gtrsim18$~keV gas , while the kSZ effect map reveals gas with $\vz\approx3000$~km/s. Figures are from \cite{Adam2017ksz}.
    } \label{fig:ksz}
\end{SCfigure}

\smallskip
\noindent
{\bf Constraining Kinematics of Warm-Hot Gas with tSZ and kSZ effects:}
Modern cosmological hydrodynamical simulations make firm predictions for the mass accretion histories of dark matter halos and the bulk and turbulent gas motions generated by mergers and the accretion process through the hierarchical growth of structure \citep[e.g.,][]{Lau2009,Vazza2009,Battaglia2012a,Zhuravleva2013,Nelson2014}.
Deep tSZ effect observations of merging systems can help address several outstanding questions in ICM, IGrM and CGM physics. Resolved measurements of the tSZ effect are well suited to probing shocks (pressure discontinuities) and distinguishing the nature of fluctuations in the ICM (adiabatic vs.\ isobaric). This has been demonstrated in deep, targeted measurements over the past decade, such as those from ALMA/ACA \citep{Basu2016,Kitayama2016,Ueda2018,DiMascolo2018}, MUSTANG \citep{Mason2010,Korngut2011,Mroczkowski2012,Young2015,Romero2015}, GISMO \cite{Mroczkowski2015}, NIKA \cite{Adam2015,Adam2017,Ruppin17a,Ruppin2018}, as well as several {\it Planck} observations of very nearby systems \citep{Eckert2013,Planck2013_X,Hurier2019}.
An example from \cite{Adam2017ksz} of NIKA's measurements of the tSZ effect substructure in a spectacular merging system is shown in Fig.~\ref{fig:ksz}, where the most significant SZ decrement reveals the hottest cluster gas, as confirmed by X-ray spectroscopy \cite{Mroczkowski2012,Adam2017ksz}. 

\vspace{-4mm}
\vspace{-4mm}

\smallskip
\noindent
{\bf Detecting Cluster Mergers and Bulk Motions with kSZ effect:}
The ability of kSZ effect to uniquely measure the peculiar velocity of a system with respect to the CMB (i.e.\ the most universal reference frame) provides a complementary view on mergers driven by the growth of structure.  This is well illustrated in the Bolocam and NIKA constraints for MACSJ0717.5+3745, which exhibits a line-of-sight velocity of $\vz\approx3000$~km/s for one particular subcluster.  
Fig.~\ref{fig:ksz} (right) shows the highest resolution kSZ effect measurement to date, using NIKA on the IRAM 30-meter telescope.  Work has been done to extend such measurements to other systems and larger samples \citep[e.g.][]{Zemcov2010,Sayers2018}, the latter of which has placed constraints on the global RMS gas motion (at 1\arcmin\ resolution) for a sample of 10 massive clusters. Future measurements of the kSZ effect will provide unique constraints on bulk and turbulent gas motions, especially in the outskirts of high-redshift clusters. These forthcoming measurements will be particularly complementary to upcoming X-ray missions like \xrism, \athena, and \lynx\ \citep{Tashiro2018,Nandra2013,Gaskin2016}, which will focus on similar measurements in the inner and intermediate regions of galaxy clusters in the low-redshift, more local universe.

\vspace{-3mm}
\begin{SCfigure}[3][h!]
 \hspace{-5mm}
  \includegraphics[height=52mm]{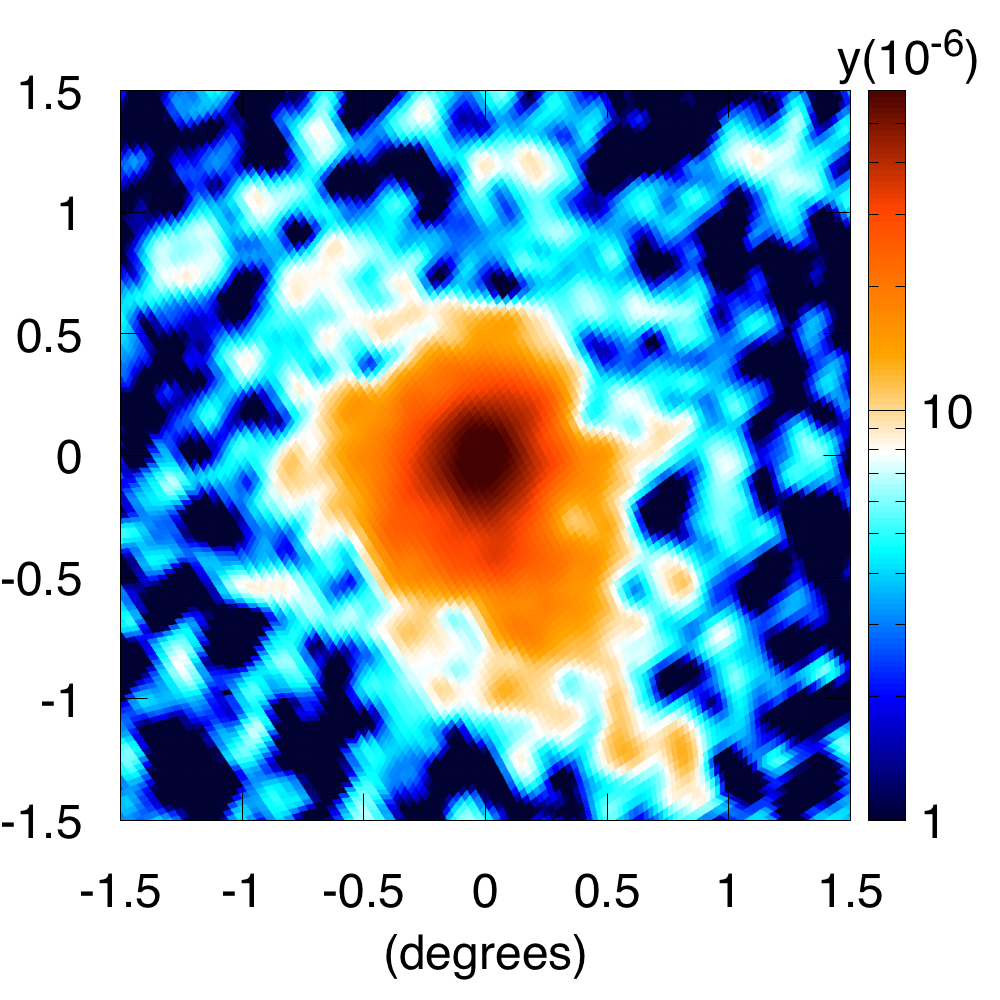}
 \hspace{-2mm}
  \includegraphics[height=52mm]{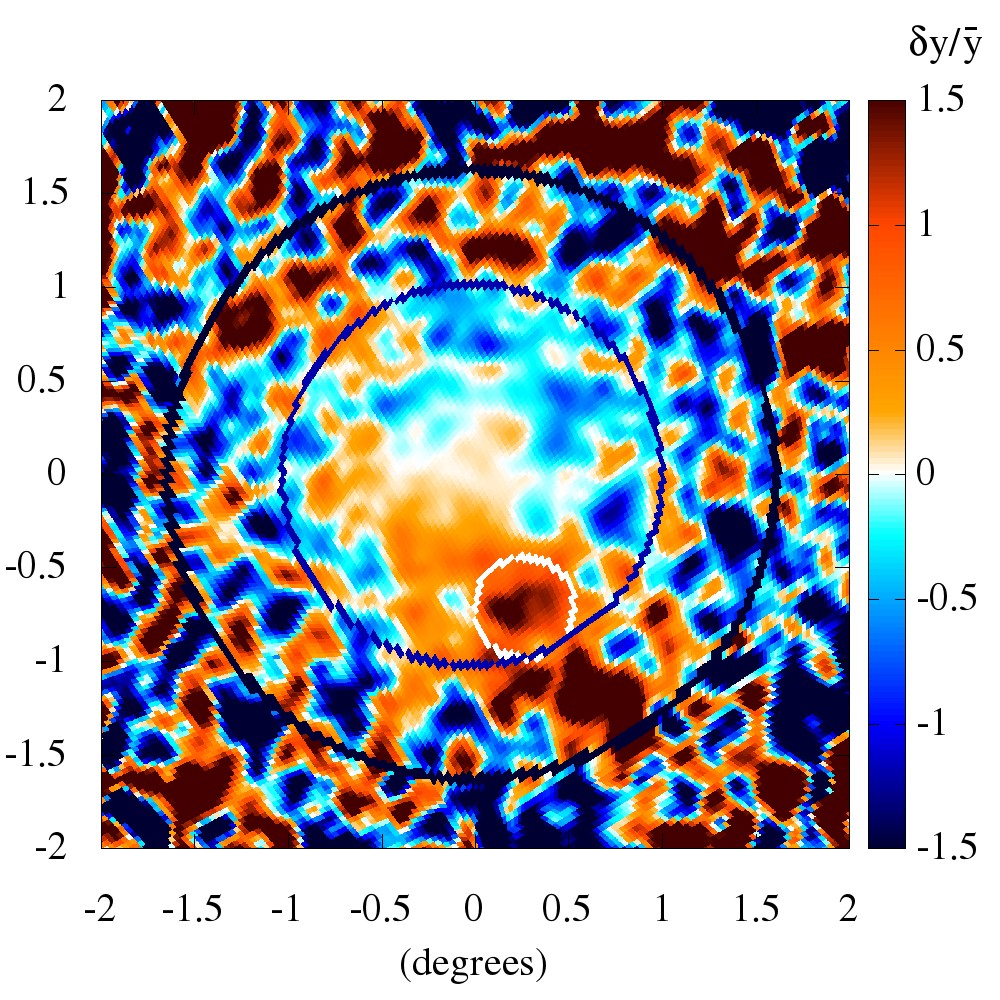}
\caption{\textit{Left:} The {\it Planck} $y$-map of the Coma cluster. \textit{Right:} Residuals after 
subtracting the average Compton $y$ profile, used to place constraints on the turbulent Mach numbers in the cluster outskirts via tSZ effect power spectral fluctuations 
\cite{Khatri2016}.}
\label{fig:coma}
\end{SCfigure}
\vspace{-4mm}

\noindent
{\bf Probing Turbulence with ICM fluctuations:}
The study of ICM fluctuations presents a relatively new tool for understanding the energetics of ICM interactions \cite{Churazov2012,Gaspari2013,Gaspari2014,Zhuravleva2014,Churazov2016}.
While usually applied to X-ray data, recent work demonstrates the potential for the application of this method to tSZ measurements, despite only probing fluctuations at the shallow level of $\left< y \right>\sim 10^{-5}$ (see Fig.~\ref{fig:coma}, right; \citep{Khatri2016}). The spectral normalization is linearly related to the ICM turbulent Mach number (and thus non-thermal pressure support). The slope of the SZ power spectrum is uniquely sensitive to thermal conduction and can determine whether density fluctuations seen in X-ray imaging are adiabatic \cite{Khatri2016} or isobaric \cite{Ueda2018,DiMascolo2018}. Further, the power spectrum of kSZ variations may give an additional constraint on turbulence, and provide key corrections to mass estimates assuming hydrostatic equilibrium \citep{Lau2013} (see white paper by Bulbul+). 

\vspace{-6mm}
\section{Prospects for the Next Decade}
\vspace{-3mm}
\bigquestion{How do we continue this rapid pace of development? How can we make a truly transformative advance in SZ observations?}
 
While theorized only shortly after the discovery of both the CMB \citep{Penzias1965} and X-ray emission from galaxy clusters \citep{Byram1966,Bradt1967}, observations of the SZ effect have only matured within the last decade to a level suitable for precise astrophysical studies.  Through the continued and rapid advances of mm/submm instrumentation, the SZ effect will strengthen its position of providing a unique window on the astrophysics of the warm and hot ionized ICM, IGrM, CGM, and WHIM and their applications to cosmology, particularly at high-$z$. 

Recently fielded and upcoming instruments with subarcminute resolution -- such as ALMA \& ACA, CCAT-p, CONCERTO, MUSTANG2, ngVLA, NIKA2, TIME, and TolTEC \cite[see][]{Crites2014,Dicker2014,Lacy2018,Lagache2018,Selina2018} -- will significantly advance the state of the art. However, most of these high-resolution instruments have modest fields of view (typically $\lesssim5'$), which will limit their ability to study the outskirt regions of galaxy clusters and other low-density environments.
This is especially true for interferometers lacking a significant single-dish component (e.g.\ ALMA\footnote{While ALMA technically includes a single-dish component, it lacks a method to modulate the signal rapidly enough to recover faint , continuum signals (e.g.\ using a chopper or nutator), and the antenna optics limit any instrumental upgrades to only a few beams at most.}, ngVLA, SKA). Similar to the 7-meter ACA component of ALMA, the inclusion of a compact array would partially mitigate this issue for the ngVLA \citep{ngvla43}. However, further ngVLA studies found that a single-dish with diameter $>45$-meters ($\gtrsim3\times$ that of the interferometric array elements) is required to access larger scales with good fidelity and high spatial dynamic range \cite{Frayer2017,ngvla54}.

Large single dish facilities covering the mm/submm bands (e.g.\ $70-500$~GHz) must therefore continue to be developed and supported.
The 100-m GBT, for instance, could host a 90~GHz tSZ camera with a 15\arcmin\ instantaneous field of view (FoV), improving both its spatial dynamic range and mapping speed. 
Similarly, a significant upgrade of the optics on the 50-m Large Millimeter Telescope (LMT) could expand its FoV to $\sim$10-15$\arcmin$, while additional improvements could be made by extending its frequency coverage beyond the 3 bands now being built for TolTEC \cite{Bryan2018}. 
Such a new generation of SZ instruments would build on the successes and lessons of the current instruments, probing scales from 10's of kpc to those nearly as large as the cluster virial radius ($>1$~Mpc) across the epoch of cluster formation ($z\lesssim 2$). 
In addition, broader spectral coverage, particularly at higher frequencies ($\gtrsim 350$~GHz), is needed for detailed velocity and
temperature studies with the kSZ and rSZ effects, and also for removing contamination from dust emission \cite{Sayers2018,Erler2018}.

A significant leap in SZ science capabilities will require
a large aperture ($\gtrsim 30$ meter), large FoV ($\gtrsim 1$ degree), multi-frequency mm/submm telescope, such as the Atacama Large Aperture Submm/mm Telescope (\href{http://atlast-telescope.org}{AtLAST}; see \cite{Bertoldi2018,deBreuck2018,Hargrave2018,Klaassen2018,Mroczkowski2018}), Large Submm Telescope (LST; see \cite{Kawabe2016}), Chajnantor Sub/millimeter Survey Telescope (CSST; see \cite{Padin2014,Golwala2018}), or CMB in High-Definition (CMB-HD; see \cite{Sehgal2019}).
With an instantaneous FoV at least 225$\times$ that of the 50-m LMT and the better atmospheric transmission of an intrinsically drier site $>5000$ meters above sea level, a mapping speed at least 500$\times$ greater can be expected for such facilities, enabling a transformative leap in sample size and serving as an invaluable complement to future X-ray facilities.

\newpage

\bibliographystyle{unsrturltrunc6}
\bibliography{references}

\end{document}